\documentclass[fleqn]{article}
\usepackage{espcrc2}

\usepackage{epsfig}

\newcommand{\ovr}{\over}
\newcommand{\til}{\tilde}

\renewcommand{\dag}{^\dagger}

\renewcommand{\>}{\rangle}
\newcommand{\gaf}{\gamma_5}

\newcommand{\al}{\alpha}
\newcommand{\be}{\beta}
\newcommand{\ga}{\gamma}
\newcommand{\de}{\delta}

\newcommand{\ze}{\zeta}

\newcommand{\ka}{\kappa}

\newcommand{\si}{\sigma}

\newcommand{\ps}{\psi}
\newcommand{\om}{\omega}

\newcommand{\beq}{\begin{equation}}
\newcommand{\eeq}{\end{equation}}
\newcommand{\bdm}{\begin{displaymath}}
\newcommand{\edm}{\end{displaymath}}
\newcommand{\bea}{\begin{eqnarray}}
\newcommand{\eea}{\end{eqnarray}}

\newcommand{\mr}{\mathrm}
\newcommand{\mb}{\mathbf}

\newcommand{\GeV}{\,\mr{GeV}}
\newcommand{\fm}{\,\mr{fm}}

\newcommand{\psb}{\bar\ps}
\newcommand{\zeb}{\bar\ze}
\newcommand{\h}{\mr{h}}
\newcommand{\psh}{\ps_\h}
\newcommand{\fA}{f_\mr{A}}
\newcommand{\fAstat}{f_\mr{A}^\mr{stat}}
\newcommand{\fAkin}{f_\mr{A}^\mr{kin}}
\newcommand{\fAspin}{f_\mr{A}^\mr{spin}}
\newcommand{\ZA}{Z_\mr{A}}

\title{
\vspace{-13mm}%
{\normalsize DESY 04-176, SFB/CPP-04-48, HU-EP-04/52
\hfill{\tt hep-lat/0409058}}\\[-2mm]
{\normalsize September 2004}\\[5mm]
Signal at subleading order in lattice HQET%
\thanks{presented by S.\ D\"urr at Lattice 04, Fermilab, USA.}%
\thanks{Work supported by DFG in framework SFB/TR-9.}%
}

\author{
S.\ D\"urr\address[DESY]{DESY, Platanenallee 6, 15738 Zeuthen,
Germany\vspace{-2mm}},
A.\ J\"uttner\address[HUB]{Humboldt Universit\"at, Newtonstrasse 15,
12489 Berlin, Germany},
J.\ Rolf\addressmark[HUB] and
R.\ Sommer\addressmark[DESY]
}

\begin{document}

\begin{abstract}
We discuss the correlators in lattice HQET that are needed to go beyond the
static theory. Based on our implementation in the Schr\"odinger functional we
focus on their signal-to-noise ratios and check that a reasonable statistical
precision can be reached in quantities like $f_{\mr{B}_\mathrm{s}}$ and
$M_\mr{B^\star}\!-\!M_\mr{B}$.
\vspace{-1mm}
\end{abstract}

\maketitle
\setcounter{footnote}{0}


\section{INTRODUCTION}

The physics of mixings and decays of B-mesons is crucial for a determination
of several entries of the CKM-matrix, e.g.\ $V_{ub}$.
To relate experimental observations to Standard Model parameters, transition
elements of the effective weak Hamiltonian must be computed in a reliable
fully non-perturbative framework, e.g.\ on the lattice.

Since the large mass of the $b$-quark evades a direct treatment at 
lattice spacings $a^{-1}\!=\!2-4\,\GeV$, effective field theory methods are
invoked.
The one tried first is the static approximation which is the leading order of
an expansion in the inverse heavy mass, i.e.\ the leading order of
HQET~\cite{Eichten:1989zv}.
Here, the heavy flavor field $\psi_\h$ satisfies
$ (1 + \gamma_0) \psi_\h = 2 \psi_\h$ and has a discretized action
\bea
S_\mr{h}^\mr{EH}\!\!&\!\!=\!\!&\!\!a^4\sum_x \psb_\h(x) D_0\psi_\h(x)
\label{eichtenhill}
\\
D_0\psi_\h(x)\!\!&\!\!=\!\!&\!\!
[\psi_\h(x)\!-\!U^\dagger(x\!-\!a\hat{\mb 0},0)\psi_\h(x\!-\!a\hat{\mb 0})]/a
\;.
\nonumber
\eea
A non-perturbative renormalization of the
effective theory --~including the subleading terms~-- is possible, i.e.\ 
divergences can be subtracted implicitly through a direct matching against
QCD in a small volume, and very precise results at short ($x_0\!<\!0.5\fm$)
distances may be obtained \cite{Heitger:2003nj}.
One of the advantages of such a strategy is that the theory maintains a
well-defined continuum limit, while this would not be
true, if the matching was performed to any fixed order in perturbation theory.
We consider this important.

There are, however, some intrinsic difficulties with this approach.
The first one is the statistical noise in correlators at large
($x_0\!>\!0.5\fm$) euclidean distances.
Brute force is no solution, since the noise-over-signal ratio grows
exponentially in $x_0$.
This is related to the $1/a$ divergence that has been subtracted.
If the HQET approach is extended beyond the leading order, $1/a^n$ divergences
are implicitly canceled, and the obvious fear is that the noise problem will
be even worse.
The second difficulty relates to the fact that any limited precision in the
matching will translate into additional uncertainties of all quantities in
the large volume; hence devising good matching conditions is an
important task.
Here, we shall address point one, i.e.\ the statistical precision that
can be reached in typical correlators at the next-to-leading order in the
HQET expansion.


\section{CORRELATION FUNCTIONS}

The Schr\"odinger functional (SF) master formula for the heavy meson decay
constant%
\footnote{For any unexplained notation we refer to
\cite{Heitger:2003nj,DellaMorte:2003mn}.}
\bdm
{F_\mr{B}\sqrt{m_\mr{B}}\ovr2L^{3/2}}\!=\!\left\{
\begin{array}{l}
\!\!\!-\ZA{\fA(T/2)\ovr\sqrt{f_1}}
\\
\!\!\!-C_\mr{PS}
(M_\mr{b}/\Lambda)\ZA^\mr{stat}{\fAstat(T/2)\ovr\sqrt{f_1^\mr{stat}}}
+\mr{O}({1\ovr m})\!\!\!\!\!
\end{array}
\right.
\edm
relates the expression in the relativistic formulation to that in the HQET
approach.
In the latter, the correlation functions at leading order are
\bea
\fAstat(x_0)\!&\!=\!&\!-{1\ovr2}\<A_0^\mr{stat}(x)O\>
\\
f_1^\mr{stat}\!&\!=\!&\!-{1\ovr2}\<O'O\>
\eea
with the static-light axial current
\beq
A_0^\mr{stat}(x)=\psb(x)\ga_0\gaf\psi_\h(x)
\eeq
and the boundary operators
\bea
O\!&\!=\!&\!\sum_{\mb{y},\mb{z}}
\zeb_\h(\mb{y})\gaf\om(\mb{y}\!-\!\mb{z})\ze(\mb{z})
\nonumber
\\
O'\!&\!=\!&\!\sum_{\mb{y}',\mb{z}'}
\zeb'(\mb{y}')\gaf\om(\mb{y}'\!-\!\mb{z}')\ze'_\h(\mb{z}')
\nonumber
\eea
on the bottom and top ($x_0\!=\!0, T$) of the SF-box.

The idea of lattice HQET at subleading order is to expand the Lagrangian
(we include coefficients for subsequent renormalization)
\bdm
L_\mr{HQET}=L_\mr{stat}
-{\til\om^\mr{kin}\ovr2m}\psb_\h\mb{D}^2\psh
-{\til\om^\mr{spin}\ovr2m}\psb_\h\mb{\si\!\cdot\!B}\psh
\edm
in the exponent and to treat the new terms as insertions in any correlator.
Expanding consistently means that one keeps just terms in $1/m$, no products
$\mr{O}(1/m^2)$.
The correlator $\fAstat$ is thus augmented by $\til\om^\mr{kin/spin}/(2m)$
times
\beq
\fA^\mr{kin/spin}(x_0)\!=\!
-{1\ovr2}\<A_0^\mr{stat}(x)\sum_u X^\mr{kin/spin}(u)O\>
\eeq
with $X^\mr{kin}(u)\!=\!\psb_\h(u)\mb{D}^2\psh(u)$ as bulk insertion or
$X^\mr{spin}(u)\!=\!\psb_\h(u)\mb{\si\!\cdot\!B}\psh(u)$, and ditto for
$f_1^\mr{stat}$.
The dimension 5 pieces of the Lagrangian appear only as insertions, and this is
crucial for the renormalizability [to order $1/m$] of the theory.

The expansion of $\fA$ to order $1/m$ reads
\bdm
\fA\!\propto\!\fAstat
\Big\{
1\!+\!{\al^{(1)}\ovr\al^{(0)}}\!{\de\!\fAstat\ovr\fAstat}
\!+\!\om^\mr{kin}{\fAkin\ovr\fAstat}
\!+\!\om^\mr{spin}{\fAspin\ovr\fAstat}
\Big\}\!
\edm
where $\om^\mr{kin}\!=\!\til\om^\mr{kin}/(2m)$ ($=\!{1\ovr2}\om_2^{(1)}$ in the
notation of\,\cite{Heitger:2003nj}) and similarly for $\om^\mr{spin}$.
An analogous expression (without the $\de\!\fAstat$ piece defined
in\,\cite{DellaMorte:2003mn}) replaces $f_1^\mr{stat}$.
The coefficients $\al^{(0/1)}, \om^\mr{kin/spin}$ are functions of
$M_ba, g_0^2$ and may be determined as in\,\cite{Heitger:2003nj}.
Hence
\bea
{F_\mr{B}\sqrt{m_\mr{B}}\ovr2L^{3/2}}\!&\!=\!&\!
-{\al^{(0)}\!\fAstat\ovr\sqrt{f_1^\mr{stat}}}
\Big\{1+{\al^{(1)}\ovr\al^{(0)}}{\de\!\fAstat\ovr\fAstat}+
\label{fbsfactor}
\\
\!&\!{}\!&\!
\om^\mr{kin}R^\mr{kin}(T/2)+\om^\mr{spin}R^\mr{spin}(T/2)
\Big\}
\nonumber
\eea
where
\bdm
R^\mr{kin}={\fAkin\ovr\fAstat}\!-\!{1\ovr2}{f_1^\mr{kin}\ovr f_1^\mr{stat}}
\,,\;
R^\mr{spin}={\fAspin\ovr\fAstat}\!-\!{1\ovr2}{f_1^\mr{spin}\ovr f_1^\mr{stat}}
\;.
\edm

In the same way, an effective mass
$m_\mr{eff}(x_0)\!=\!a^{-1}\log(\fA(x_0)/\fA(x_0\!+\!a))$ is $1/m$-expanded as
\bea
m_\mr{eff}(x_0)\!&\!=\!&\!m_\mr{eff}^\mr{stat}(x_0)+\de m+
\nonumber
\\
\!&\!{}\!&\!\om^\mr{kin}r^\mr{kin}(x_0)+\om^\mr{spin}r^\mr{spin}(x_0)
\label{neweffmass}
\eea
with 
$m_\mr{eff}^\mr{stat}(x_0)$ from $\fAstat(x_0)$ and
\bea
r^\mr{kin}(x_0)\!&\!=\!&\!a^{-1}
\Big[
{\fAkin(x_0)\ovr\fAstat(x_0)}-{\fAkin(x_0\!+\!a)\ovr\fAstat(x_0\!+\!a)}
\Big]
\nonumber
\\
r^\mr{spin}(x_0)\!&\!=\!&\!a^{-1}
\Big[
{\fAspin(x_0)\ovr\fAstat(x_0)}-{\fAspin(x_0\!+\!a)\ovr\fAstat(x_0\!+\!a)}
\Big]
\nonumber
\;.
\eea
Here, the contribution ${\al^{(1)}\ovr\al^{(0)}}
[{\de\!\fAstat(x_0)\ovr\fAstat(x_0)}
-{\de\!\fAstat(x_0\!+\!a)\ovr\fAstat(x_0\!+\!a)}]$
has been suppressed, since in the plateau region the two terms would cancel,
while $r^\mr{kin/spin}$ approach a constant value each.


\section{NUMERICAL RESULTS}


\begin{figure}[b!]
\vspace{-7mm}
\hspace{-2mm}
\includegraphics[width=76mm,angle=0]{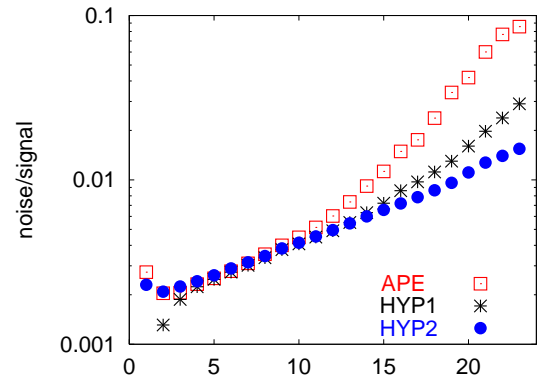}
\vspace{-14mm}
\caption{Noise-over-signal ratio for the heavy-light correlator $\fA^\mr{kin}$,
using three discretizations of the heavy propagator. Similar graphs are obtained
at leading order in the $1/m$ expansion~\cite{DellaMorte:2003mn}.}
\label{fig:ns_fAkin}
\end{figure}

The problem of large fluctuations in the static-light propagator was solved by
modifying the covariant derivative in (\ref{eichtenhill}).
As shown in \cite{DellaMorte:2003mn}, an APE- or HYP-smeared link
$V\dag(x\!-\!a\hat\mb{0},0)$ instead of $U\dag$ leads to an improvement in the
noise-over-signal ratio of $\fAstat(x_0)$ which grows exponentially with $x_0$
while the {\em discretization errors remain at the same level\/}.
Fig.\ \ref{fig:ns_fAkin} shows that a similar improvement is found at
subleading order in the $1/m$ expansion.
HYP-links \cite{Hasenfratz:2001hp} with $(\al_1,\al_2,\al_3)\!=\!(1.0,1.0,0.5)$
do better than those with $(\al_1,\al_2,\al_3)\!=\!(0.75,0.6,0.3)$, which, in
turn, are better than APE-links \cite{Albanese:1987ds} with staple-only
contribution and no $SU(3)$-projection.
The same ordering was found at leading order.

At this point we cannot compute the subleading contribution to $f_\mr{B_s}$,
since the matching coefficients $\al^{(0/1)}, \om^\mr{kin/spin}$ have not yet
been determined, but we can assess the statistical error of such a contribution.
Our figure and numbers stem from a quenched simulation in a $16^3\!\times\!24$
SF-box at $\be\!=\!6.0$ with $\ka\!\simeq\!\ka_\mr{strange}$ and 5600
measurements, and we shall quote only the results for the improved
HYP-parametrization.
For the two correction terms in (\ref{fbsfactor}) we find the values
$aR^\mr{kin} \!=\!-0.78(14)$, 
$aR^\mr{spin}\!=\!-0.83(03)$. 

For an interpretation of their statistical errors, we need a rough figure for
the coefficients $\om^\mr{kin/spin}$.
Up to typical (logarithmically $a$-dependent) renormalization factors of order
one, they can be estimated by their tree-level values
$a^{-1}\om^\mr{kin/spin}\!\simeq\!1/(2am)\!\simeq\!0.2$
(from $a^{-1}\!\simeq\!2\GeV$ and $m\!\simeq\!5\GeV$).
In addition, the renormalization of $R^\mr{kin/spin}$ does require power
divergent subtractions, but they are contained in $\alpha^{(0)}$
(for which, therefore, a tree-level estimate would be no good).

From $\om^\mr{kin/spin}\!\simeq\!0.2$ one gets a statistical error of the
kinetic contribution of 3\% and an error of the spin contribution of 0.6\%
in expression (\ref{fbsfactor}).
Neglecting uncertainties in the $\al^{(0/1)}$ and $\om^\mr{kin/spin}$, it
appears that already with the methods applied here, one can compute
$f_\mr{B_s}$ at subleading order in the HQET expansion in such a way that the
noise in the dimension 5 correlators increases the error by $\sim$3-4
percentage points.

A simpler application, where even a first estimate may be given, is the
vector-pseudoscalar splitting $M_\mr{B^\star}\!-\!M_\mr{B}$.
This effect sets in at subleading order in the HQET expansion,
\beq
M_\mr{B^\star}-M_\mr{B}\!=\!{4\ovr3}\om^\mr{spin}r^\mr{spin}(x_0)
\eeq
and has first been estimated in a lattice computation in
\cite{Bochicchio:1991cy}.
Our result for the large-$x_0$ asymptotic is $a^2r^\mr{spin}\!=\!0.060(5)$.
With $a^{-1}\om^\mr{spin}\!\simeq\!0.2$ one gets
$M_\mr{B^\star}\!-\!M_\mr{B}\!=\!0.016(1)(2\!\GeV)\!=\!0.032(3)\GeV$, which
should be compared to the experimental result $0.046\GeV$.
Hence, the lattice value falls short by about a third.
Whether this reflects a large renormalization factor (similar to $c_\mr{SW}$),
a slow convergence of the HQET expansion or a genuine quenching artefact
remains to be seen,
although the results of \cite{Heitger:2004gb} render the second
explanation rather unlikely. 
We emphasize that these figures are just indicative; reliable numbers can be
given only when $\om^\mr{spin}$ has been determined and smaller lattice
spacings have been considered.


\section{CONCLUSION}


We have attempted a first test with correlator ratios needed to compute
observables in the heavy-light system beyond the static approximation.
The natural fear that they will, for large euclidean distances, be even noisier
than the leading order piece seems not to become true.
This is because the reduction of the noise-over-signal ratio via a better
HQ-discretization works more efficiently at subleading order.
The statistical precision may be further improved, e.g.\ by employing a smaller
time extent in $f_1^\mr{stat}$ and $f_1^\mr{kin/spin}$ (together with a careful
choice of the wave function, see \cite{DellaMorte:2003mn}) or by the
methods of \cite{Green:2003zz}.
We conclude that lattice HQET at subleading order looks promising
enough to motivate further studies.

\vspace{2mm}

We thank A.\,Shindler and M.\,Della Morte for checks on the implementation and
discussions.


\end{document}